\documentclass[preprint,showpacs,preprintnumbers,amsmath,amssymb,superscriptaddress]{revtex4}

\usepackage{amssymb}
\usepackage{graphicx}
\usepackage{dcolumn}
\usepackage{bm}
\usepackage[CJKbookmarks=true, dvipdfm, colorlinks, linkcolor=blue, anchorcolor=black, citecolor=blue, urlcolor=blue]{hyperref}

\begin{document}


\title{Charge fluctuations and nodeless superconductivity in quasi-one-dimensional Ta$_{4}$Pd$_{3}$Te$_{16}$ revealed by $^{125}$Te-NMR and $^{181}$Ta-NQR} 

\author{Z. Li}
\affiliation{Institute of Physics and Beijing National Laboratory for Condensed Matter Physics, Chinese Academy of Sciences, Beijing 100190, China}
\author{W. H. Jiao}
\affiliation{Department of Physics, State Key Lab of Silicon Materials and Center of Electron Microscope, Zhejiang University, Hangzhou 310027, China}
\author{G. H. Cao}
\affiliation{Department of Physics, State Key Lab of Silicon Materials and Center of Electron Microscope, Zhejiang University, Hangzhou 310027, China}
\affiliation{Collaborative Innovation Center of Advanced Microstructures, Nanjing 210093, China}
\author{Guo-qing Zheng}%
\affiliation{Institute of Physics and Beijing National Laboratory for Condensed Matter Physics, Chinese Academy of Sciences, Beijing 100190, China}
\affiliation{Department of Physics, Okayama University, Okayama 700-8530, Japan}

\date{\today}

\begin{abstract}
We report $^{125}$Te nuclear magnetic resonance and $^{181}$Ta nuclear quadrupole resonance studies on  single-crystal Ta$_{4}$Pd$_{3}$Te$_{16}$, which has a quasi-one-dimensional structure and superconducts below $T_{\rm c}=4.3$ K. $^{181}$Ta with spin $I=7/2$ is sensitive to quadrupole interactions, while $^{125}$Te with spin $I=1/2$ can only relax by magnetic interactions. By comparing the spin-lattice relaxation rate ( $1/T_{1}$) of $^{181}$Ta and  $^{125}$Te, we found that electric-field-gradient (EFG) fluctuations develop below $80$ K. The EFG fluctuations are enhanced with decreasing temperature due to the fluctuations of a charge density wave  that sets in at $T_{\rm CDW}=20$ K, below which the spectra are broadened and $1/T_{1}T$ drops sharply. In the superconducting state, $1/T_{1}$ shows a Hebel-Slichter coherence peak just below $T_{\rm c}$ for $^{125}$Te, indicating that Ta$_{4}$Pd$_{3}$Te$_{16}$ is a full-gap superconductor without nodes in the gap function. The coherence peak is absent in the $1/T_{1}$ of $^{181}$Ta due to the strong EFG fluctuations.

\end{abstract}

%
\maketitle

\section{\label{sec:Introduction}Introduction}
Low-dimensional systems often show rich physical phenomena, such as charge density wave (CDW), spin density wave (SDW) and superconductivity. When superconductivity coexists with or adjoins another ordered state, it usually has an unconventional nature. The high transition temperature ($T_{\rm c}$) superconductivity has been realized in two-dimensional copper-oxides\cite{Bednorz1986} and iron pnictides\cite{KamihalaY2008}, where the superconducting phase is located in the vicinity of an antiferromagnetic phase. This raises an interest in finding unconventional superconductivity in low-dimensional materials. Quasi one-dimensional Li$_{0.9}$Mo$_{6}$O$_{17}$\cite{GreenblattM} and organic compounds including (TMTSF)$_{2}$PF$_{6}$ \cite{JeromeD} and (TMTSF)$_{2}$ClO$_{4}$ \cite{BechgaardK} were discovered to be  superconductors with possible unconventional natures. One-dimensional systems also attract intensive  theoretical interest for their simplicity\cite{CarrST2002}.

Ta$_{4}$Pd$_{3}$Te$_{16}$ with a quasi-one-dimensional crystal structure was discovered to be superconducting with $T_{\rm c}\sim 4.6$ K \cite{Jiao2014}. It consists of PdTe$_{2}$ chains, TaTe$_{3}$ chains, and Ta$_{2}$Te$_{4}$ double chains along the crystallographic $b$-axis\cite{MarA1991}, as shown in Fig. \ref{fig:FigStr}.
Band structure calculations indicate that the density of states (DOS) at the Fermi level are mainly derived from Te $p$-orbitals\cite{Singh2014}. Although scanning tunneling microscopy (STM) found that the superconducting gap structure in this system is more likely anisotropic without nodes\cite{DuZY2015,FanQ2015},  nodal gap behaviors were claimed by thermal conductivity\cite{PanJ2015} and specific heat measurements\cite{JiaoWH2015}. STM study suggested that the system is in the vicinity of an ordered state that shows a periodic modulation\cite{FanQ2015}. This suggestion is consistent with the observation that the magnetoresistance shows an $H$-linear behavior without saturation up to $50$ T\cite{XuXF2015}. However, the nature of the ordered state is unclear and the transition temperature is still undetermined.

In this paper, we report nuclear magnetic resonance (NMR) and nuclear quadrupole resonance (NQR) investigations on Ta$_{4}$Pd$_{3}$Te$_{16}$. The spin-lattice relaxation rate ($1/T_{1}$) divided by the temperature, $1/T_{1}T$, of $^{181}$Ta increases dramatically below $80$ K, but $1/T_{1}T$ of $^{125}$Te is almost a constant. These results indicate strong electric-field-gradient (EFG) fluctuations, since $^{181}$Ta has a nuclear spin $I=7/2$ with a large nuclear quadrupole moment that couples to EFG, but $^{125}$Te does not. Upon cooling, both the FWHM of the spectra and $1/T_{1}T$ show a sudden change at $T=20$ K, indicating a CDW transition takes place at $T_{\rm CDW}=20$ K. In the superconducting state, a Hebel-Slichter coherence peak appears in the temperature dependence of $1/T_{1}$ of $^{125}$Te just below $T_{\rm c}$, which indicates that Ta$_{4}$Pd$_{3}$Te$_{16}$ is a  fully-gapped superconductor. Strong EFG fluctuations eliminate the coherence peak in the temperature dependence of $1/T_{1}$ of $^{181}$Ta.

\section{\label{sec:Methods}Experimental details}
The Ta$_{4}$Pd$_{3}$Te$_{16}$ single crystals were prepared by the self-flux method, as reported in Ref.\onlinecite{Jiao2014}. Superconductivity with $T_{\rm c} \sim 4.3$ K is confirmed by dc magnetization measured by a superconducting quantum interference device (SQUID). For $^{125}$Te NMR measurements, the $b$ axis of the single crystals were aligned to the magnetic field, $H \parallel b$. The $^{125}$Te nucleus has a 
nuclear gyromagnetic ratio $\gamma = 13.454$ MHz/T. For $^{181}$Ta NQR measurements, crystals were crushed into powders. The $^{181}$Ta nucleus has a 
nuclear gyromagnetic ratio $\gamma = 5.096$ MHz/T.
The NMR and NQR measurements were carried out by using a phase coherent spectrometer. The $^{125}$Te NMR spectra were obtained by scanning radio frequency at fixed magnetic fields and also checked by scanning the magnetic field at fixed frequencies.
The $^{181}$Ta spectra were obtained by scanning the radio frequency at zero magnetic field. The $1/T_{\rm 1}$ was measured by using a single saturation pulse.
The recovery curve of the $^{125}$Te nuclear magnetization has a simple exponential form, $1-M(t)/M(\infty)=$exp$(-t/T_{1})$, where $M(t)$ is the nuclear magnetization at a time $t$ after the saturation pulse. The nuclear magnetization curve of $^{181}$Ta was well fitted to $1-M(t)/M(\infty)=0.04{\rm exp}(-3t/T_{1})+0.37{\rm exp}(-9.3t/T_{1})+0.59{\rm exp}(-18.3t/T_{1})$ for an asymmetry parameter $\eta =0.29$ (see later) \cite{Chepin1991}.

\section{\label{sec:ResultDiscussions}Results and Discussions}
\subsection{\label{sec:Normal}Normal state}

\begin{figure}
\includegraphics[width=7.5cm,clip]{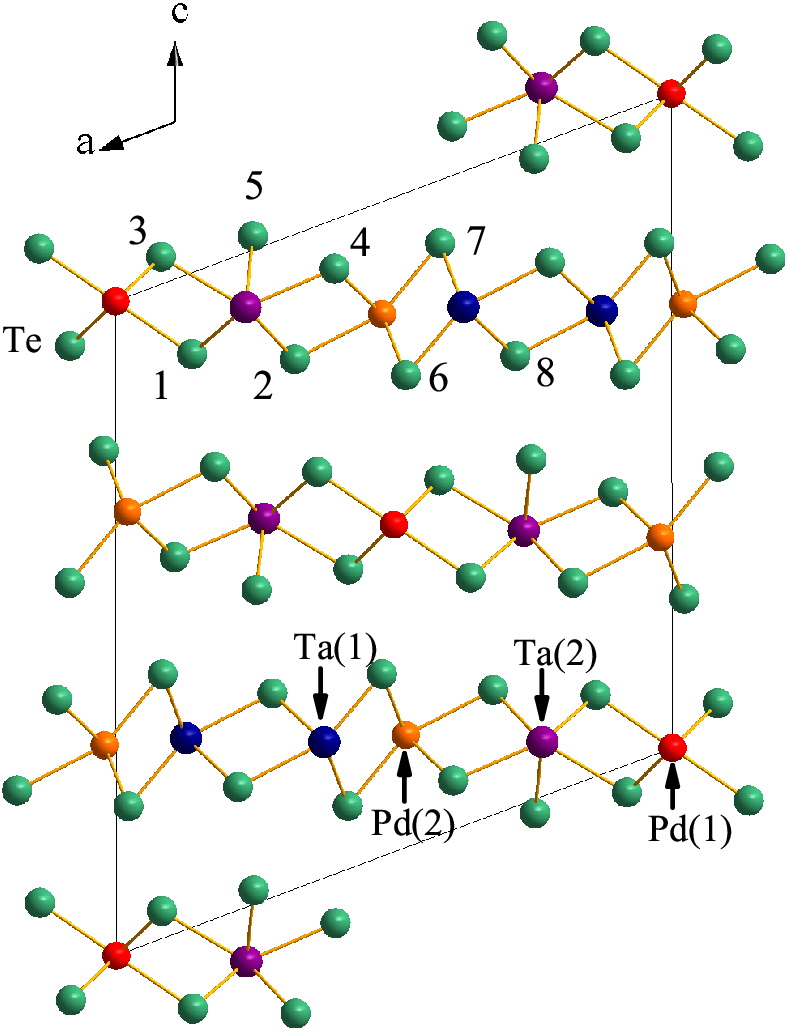}
\caption{\label{fig:FigStr} (Color online) Structure of Ta$_{4}$Pd$_{3}$Te$_{16}$ viewed from the $b$-axis direction. There are two different sites of Ta and two different sites of Pd. Eight sites of Te are labeled by numbers.}
\end{figure}

\begin{figure}
\includegraphics[width=8.6cm,clip]{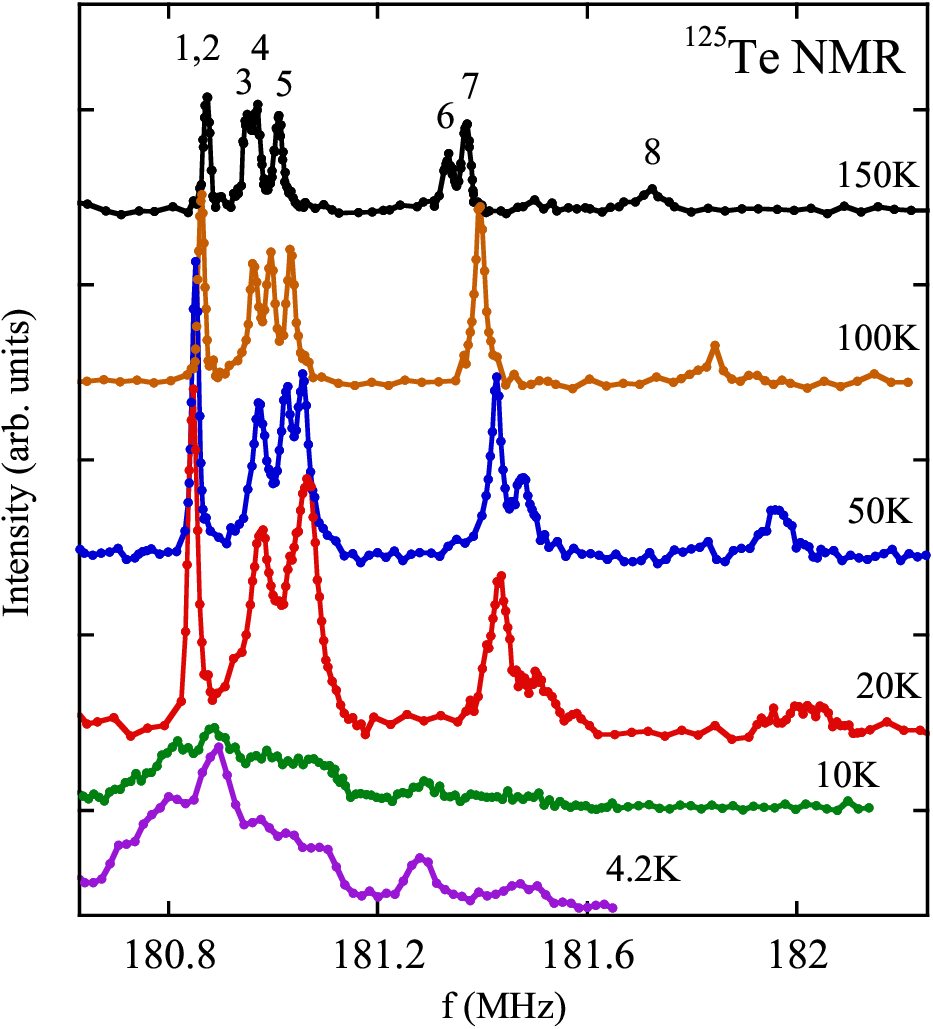}
\caption{\label{fig:TeSpectra} (Color online) $^{125}$Te NMR spectra with $\mu_{0}H(=13.43{\rm T}) \parallel b$. The vertical axes were offset for clarity. Below $20$ K, the signal is too weak to be detected above $181.6 {\rm MHz}$.}
\end{figure}

\begin{figure}
\includegraphics[width=7.5cm,clip]{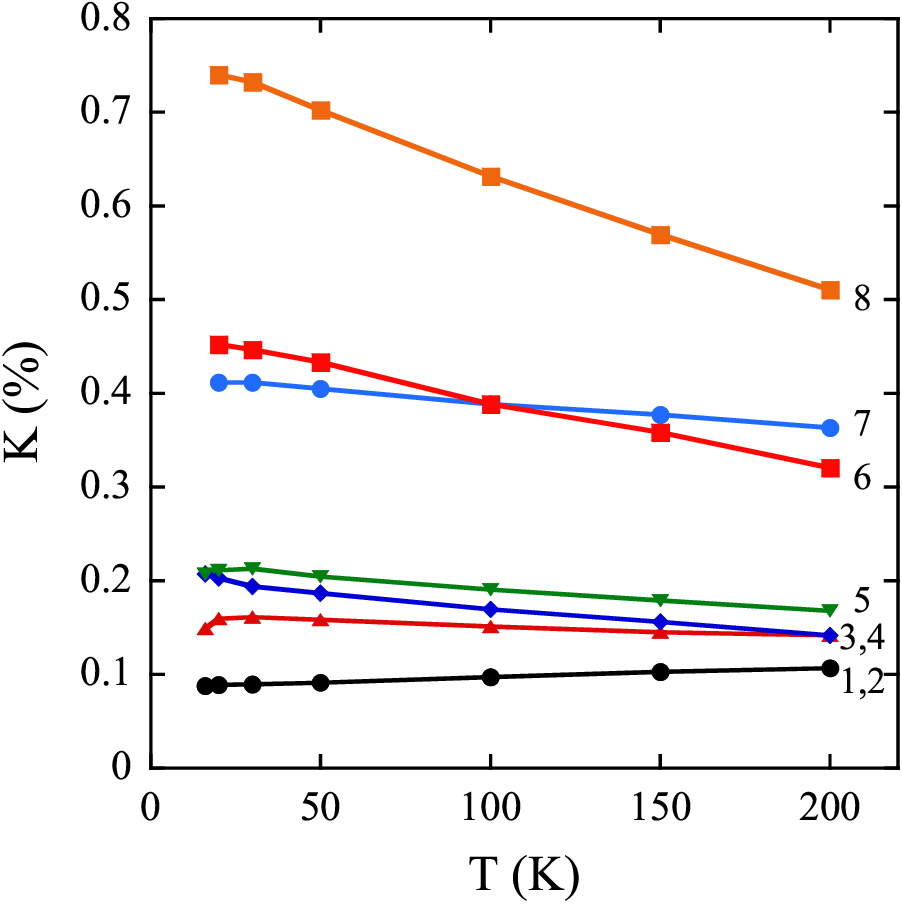}
\caption{\label{fig:TeK} (Color online) Temperature dependence of Knight shift of $^{125}$Te. The numbers mark the peaks corresponding to Fig. \ref{fig:TeSpectra}}
\end{figure}

\begin{figure}
\includegraphics[width=5cm,clip]{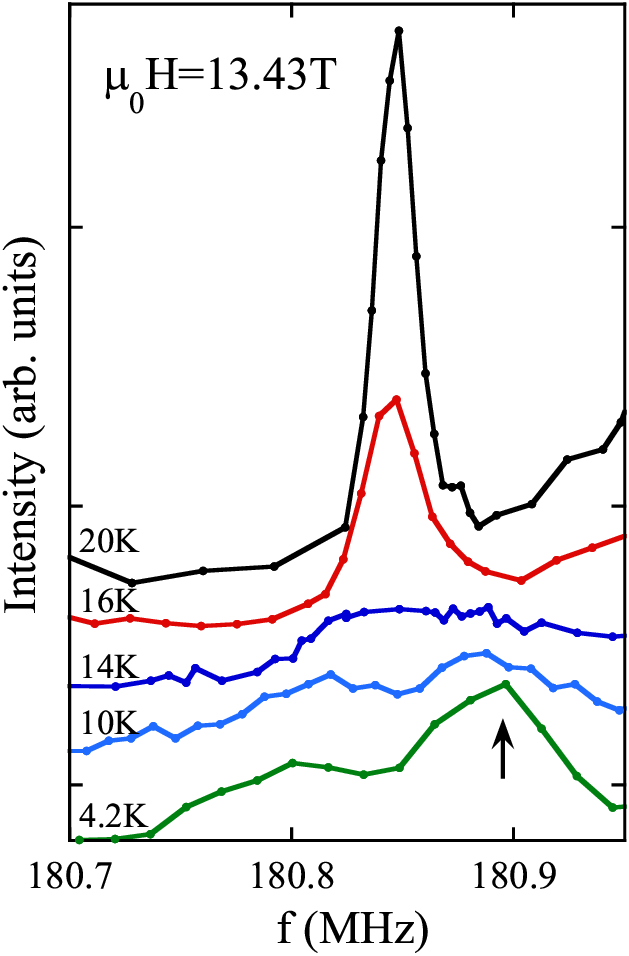}
\caption{\label{fig:TePeak1} (Color online) Peaks 1 and 2 of $^{125}$Te NMR spectra with $H \parallel b$. The vertical axes were offset for clarity. Below $20$ K, peaks $1$ and $2$ lose intensity and  split gradually. $T_{1}$ was measured at the peak $2$ marked by the arrow.}
\end{figure}

Figure \ref{fig:TeSpectra} shows the frequency-scanned  $^{125}$Te NMR spectra with $\mu_{0}H(=13.43{\rm T}) \parallel b$ axis at different temperatures. There are eight Te sites in the unit cell as shown in Fig. \ref{fig:FigStr}, so one should see eight resonance peaks. The Knight shift of these peaks is shown in Fig. \ref{fig:TeK}. Peaks $1$ and $2$ have the same frequency above $20$ K, but split at low temperature due to a CDW transition, which will be discussed later in detail. The evolution of the splitting is shown in Fig. \ref{fig:TePeak1}.

\begin{figure}
\includegraphics[width=8.6cm,clip]{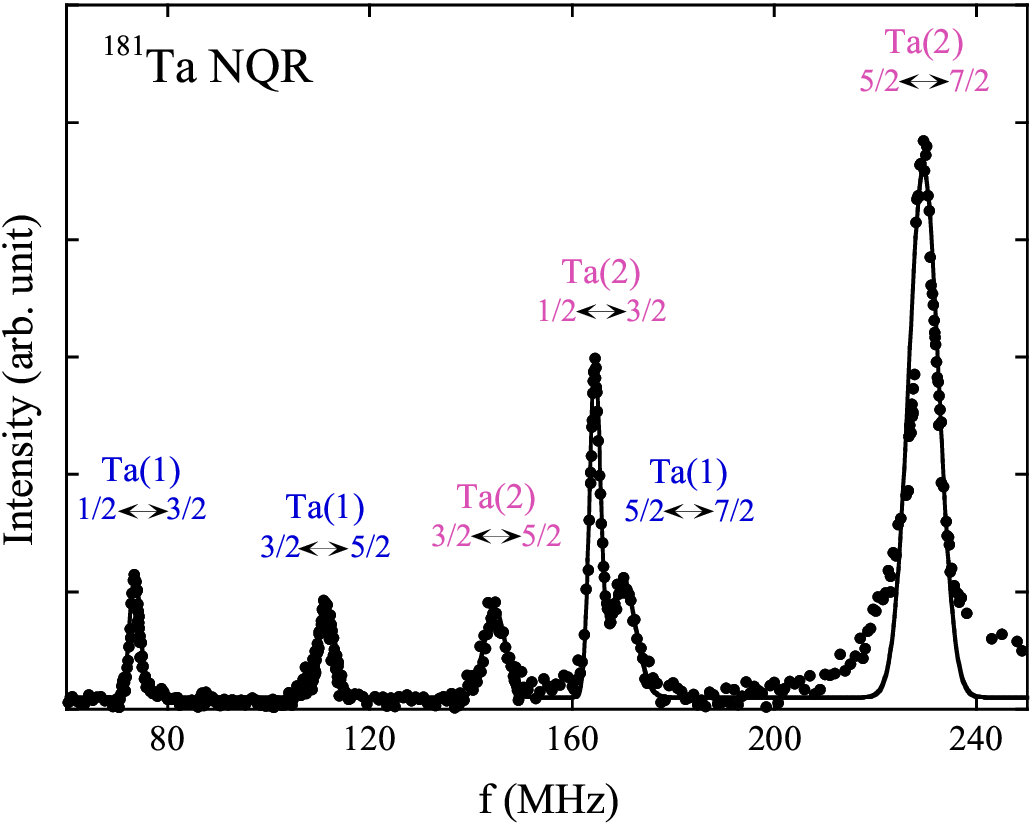}
\caption{\label{fig:TaSpectra} (Color online) $^{181}$Ta NQR spectra at $30$K. Curves are Gaussian fits.}
\end{figure}

Figure \ref{fig:TaSpectra} shows the frequency-scanned $^{181}$Ta NQR spectra at $30$ K. $^{181}$Ta has $I=7/2$ which will result in three NQR peaks. In Ta$_{4}$Pd$_{3}$Te$_{16}$, Ta has two different sites, as shown in Fig. \ref{fig:FigStr}, which will double the resonance peaks. The six peaks seen in Fig. \ref{fig:TaSpectra} can be categorized into two sets, each of which have three peaks. The NQR spectra arise from the interaction between the nuclear quadrupole moment $Q$ and the EFG $V_{\alpha\alpha}=\frac{\partial ^{2} V}{\partial \alpha ^{2}}$ $(\alpha=x,y,z)$, where $V$ is the electrical potential. The Hamiltonian is \cite{SlichterCP}
\begin{eqnarray}\label{Eq:nuQ}
\mathcal{H}_{Q}=\frac{eQV_{zz}}{4I(2I-1)}\left[ \left( 3I_{z}^{2}-I^{2} \right)+\frac{\eta}{2} \left( I_{+}^{2}+I_{-}^{2} \right) \right],
\end{eqnarray}
where $\eta=\frac{V_{xx}-V_{yy}}{V_{zz}}$ is the asymmetry parameter of the EFG. $^{181}$Ta possesses a large nuclear quadrupole moment, $Q=4.2\times 10^{-24}$ cm$^{2}$, which is ten times larger than those for such nuclei as $^{75}$As or $^{63}$Cu. This is why the NQR frequency observed is much higher than that of $^{75}$As in iron-arsenides \cite{Kawasaki2015} and Cr-arsenides \cite{RbCrAs} or $^{63}$Cu in cuprates \cite{ZhengGQ1996} where the  $^{75}$As-NQR frequency  is less than 50 MHz and the $^{63}$Cu-NQR frequency  is less than 40 MHz.

\begin{figure}
\includegraphics[width=6cm,clip]{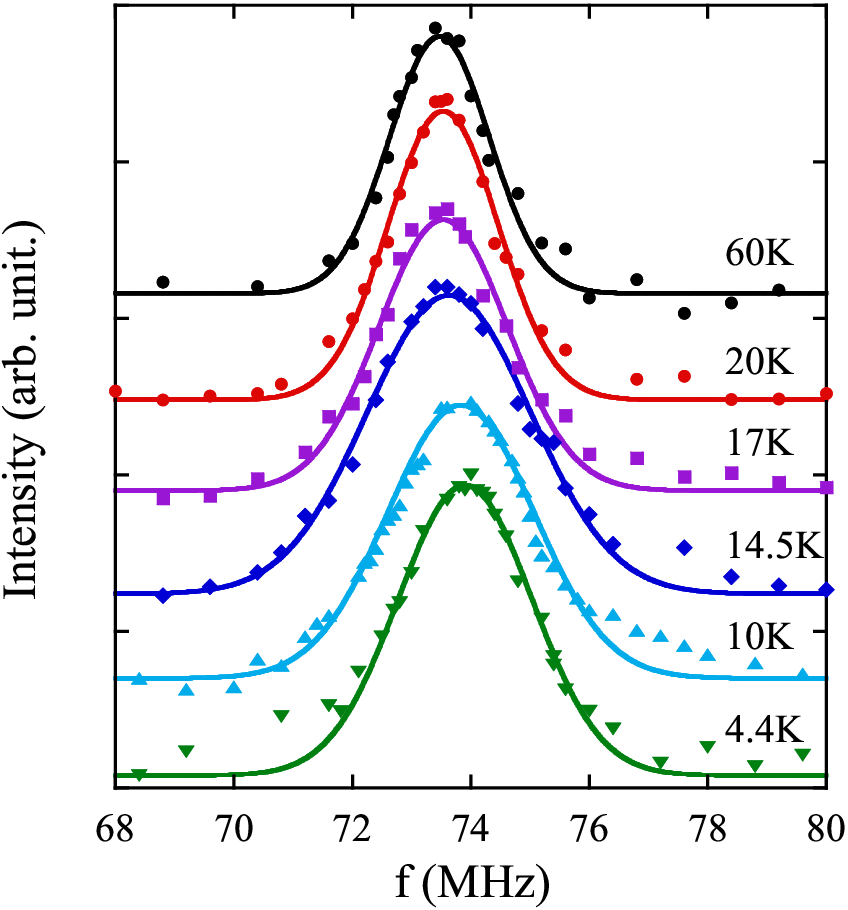}
\caption{\label{fig:FigTapeak1} (Color online) $^{181}$Ta$(1)$ NQR peak ($m=\pm1/2\longleftrightarrow\pm3/2$ transition) at some typical temperatures. }
\end{figure}

In the case of $\eta=0$, the resonance frequency from the $m \leftrightarrow m-1$ transition is $\nu_{m \leftrightarrow m-1}=(m-\frac{1}{2})\frac{3eQV_{zz}}{2I(2I-1)}$, and three transition lines will be equally separated. In the case of finite $\eta$, however, the three lines are not equally spaced, and the transition frequency of $\pm 1/2 \longleftrightarrow \pm 3/2$ can even be higher than that of  $\pm 3/2 \longleftrightarrow \pm 5/2$ for $\eta \geq$ 0.6 \cite{Yang}.
Obviously, Ta$_{4}$Pd$_{3}$Te$_{16}$ is not axial symmetric at the Ta site, as can be seen in Fig. \ref{fig:FigStr}, which can result in an asymmetric EFG.
Indeed, the obtained spectrum can be fitted by using $\eta_{1}=0.29$ for Ta$(1)$ and $\eta_{2}=0.7$ for Ta$(2)$, respectively. It is the large $\eta$ at Ta$(2)$ that make the transition of $\pm 1/2 \longleftrightarrow \pm 3/2$ a higher frequency than that of $\pm 3/2 \longleftrightarrow \pm 5/2$.
This is the unique solution of the site assignment, due to some severe restrictions such as that  the frequency of the $\pm 3/2 \longleftrightarrow \pm 5/2$ transition cannot be larger than twice of the frequency of the $\pm 1/2 \longleftrightarrow \pm 3/2$ transition \cite{Yang}.
At this moment, it is hard to identify which Ta site in the structure has larger $\eta$, and the labels of $(1)$ and $(2)$ in Fig. \ref{fig:TaSpectra} might not correspond to those in Fig. \ref{fig:FigStr}.

\begin{figure}
\includegraphics[width=8cm,clip]{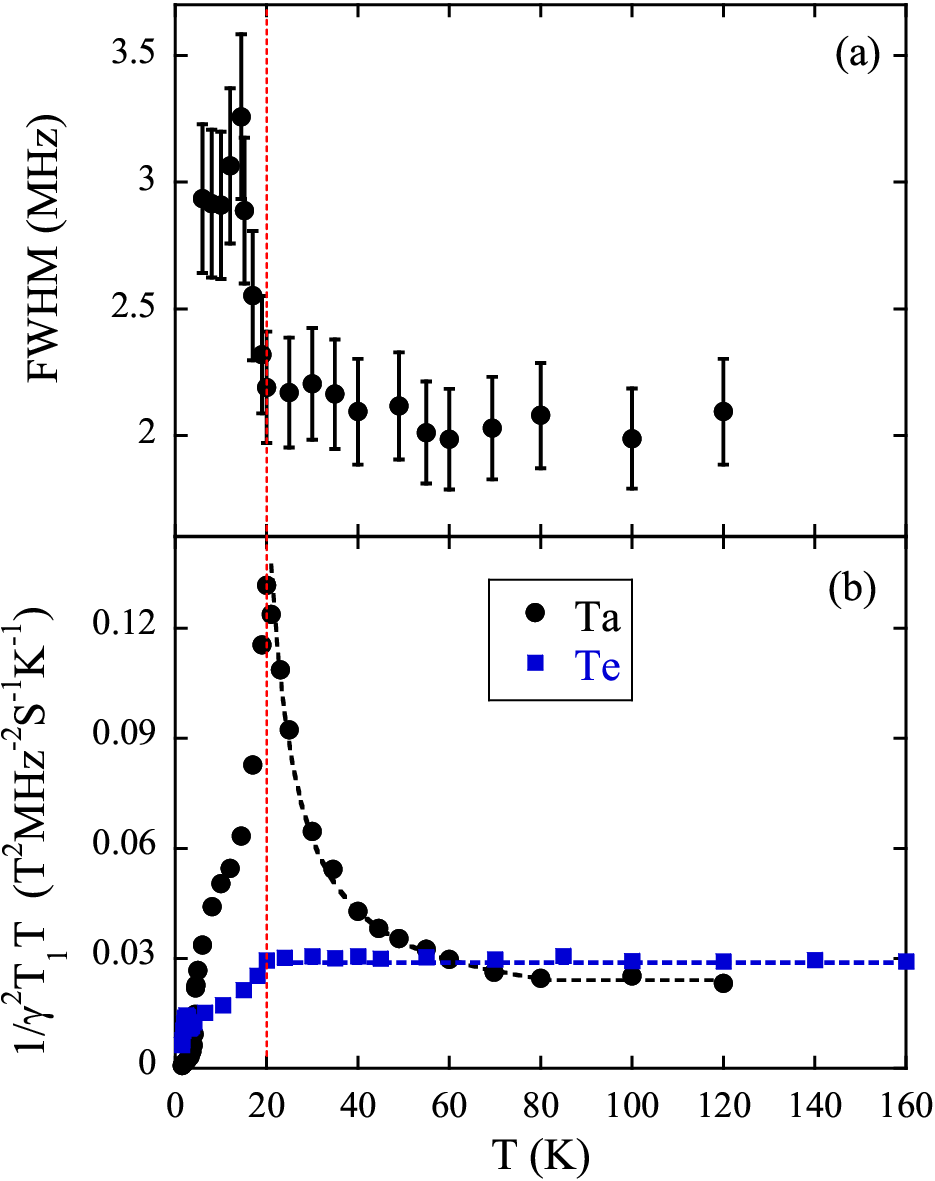}
\caption{\label{fig:Fig_fT1T} (Color online) (a) The temperature dependence of FWHM of the peak $^{181}$Ta$(1)$ ($\pm1/2\longleftrightarrow\pm3/2$ transition). (b) The temperature dependence of $1/\gamma^{2}T_{1}T$ vs $T$ for  $^{125}$Te and $^{181}$Ta. The curve and line are guides to the eyes. The red vertical line indicates the position of $T_{\rm CDW}$.}
\end{figure}

We studied the temperature dependence of the peak at around $73.5$ MHz ($\pm 1/2 \longleftrightarrow \pm 3/2$). The spectra at some typical temperatures are shown in Fig. \ref{fig:FigTapeak1}. Above $20$ K, the spectra shape is almost unchanged with changing temperature, but it becomes broad below $20$ K. The temperature dependence of the FWHM is plotted in Fig. \ref{fig:Fig_fT1T} (a). An increase at $20$ K is identified in the FWHM, which points to a phase transition. As discussed below, the spin-lattice relaxation rate result indicates that this transition is a CDW transition.

As shown in Fig. \ref{fig:Fig_fT1T} (b), in order to compare $1/T_{1}T$ of different nuclei, $1/T_{1}T$ is normalized by the square of gyromagnetic ratio $\gamma^{2}$, since $1/T_{1}T$ is proportional to $\gamma^{2}$ for magnetic relaxation.
We measured $T_{1}$ of $^{125}$Te at the peak of No. $2$. The measurement was done  at a low field of $2.308$ T, under which $T_{\rm c}=2.9$ K as measured by the ac susceptibility using an \emph{in situ} NMR coil. The  $T_{1}$  of $^{181}$Ta was measured at the NQR peak at around $73.5$ MHz ($\pm 1/2 \longleftrightarrow \pm 3/2$ transition).

$^{125}$Te with spin $I=1/2$ can only relax by
magnetic interactions, while $^{181}$Ta with spin $I=7/2$ can relax by both
magnetic and quadrupole interactions. 
By comparing $1/T_{1}T$ of the two nuclei, we can obtain the information on EFG fluctuations.
Above $20$ K, $1/\gamma^{2}T_{1}T$ of $^{125}$Te is $T$-independent and obeys the Korringa relationship as in a normal metal.
Above $80$ K, $1/\gamma^{2}T_{1}T$ of $^{181}$Ta is similar to that of $^{125}$Te, which means that magnetic interaction dominates the relaxation. Below $80$K, $1/\gamma^{2}T_{1}T$ of $^{181}$Ta increases dramatically and becomes much larger than that of $^{125}$Te, which implies a development of EFG fluctuations due to the fluctuations of the order parameter that describes the phase transition at $T=20$ K. Thus, such an order parameter is a charge density wave, whose fluctuations couple to the nuclear quadrupole moment of $^{181}$Ta.
This is consistent with the electronic structure calculation which implies that Ta$_{4}$Pd$_{3}$Te$_{16}$ prefers the CDW rather than the SDW\cite{Singh2014}. At $T_{\rm CDW}=20$ K, $1/T_{1}T$ of $^{181}$Ta shows a clear peak due to the critical slowing down of the EFG fluctuations. $1/T_{1}T$ of $^{125}$Te also decreases below $T_{\rm CDW}$, indicating a loss of density of state due to the ordered state. Inferring  EFG fluctuations by comparing a result for $I\geq$1 with that of $I$=1/2 was also used before. In LaPt$_{2-x}$Ge$_{2+x}$, moderate EFG fluctuations were derived by comparison of $1/T_{1}$ of $^{139}$La ($I$=7/2) and $^{195}$Pt ($I$=1/2) \cite{SMaeda2015}. In the iron pnictide BaFe$_{2}$(As$_{1-x}$P$_{x}$)$_{2}$, orbital fluctuations were inferred by comparing the results of $^{75}$As ($I$=3/2) and $^{31}$P ($I$=1/2) \cite{Curro2016}.
Meanwhile, it would be interesting to explore the influence of such charge fluctuations on  other properties, {\it i.e.}, transport properties, in more detail. In iron-based high temperature superconductors, quantum fluctuations of charge/orbital order are known to give rise to the unconventional normal-state properties \cite{Zhou,Fisher}, and may even be responsible for the  superconductivity.

Coming back to the results of the spectra, when the modulation period of the electric field is larger than the lattice constant due to a CDW, there are two or more atoms feeling different EFGs within one period and the spectra should split or be broadened, like in SrPt$_{2}$As$_{2}$\cite{KawasakiS2015}. In the plane wave case, however, the spatial distribution of NQR frequency has the form $\nu=\nu_{\rm Q}+\nu_{1}{\rm sin}(kx+\phi)$, where $\nu_{1}$ corresponds to the amplitude of the modulation wave and $\phi$ is a phase \cite{BlincR}. In the present case, the electric field modulation period along the $b$ axis is $2b$\cite{FanQ2015}, that is, $k=2\pi/2b$. Both Ta$(1)$ and Ta$(2)$ have the same position along the $b$ axis\cite{MarA1991}, that is, $x=nb$, where $n$ is an integer. When $\phi=0$, the sinusoidal term becomes zero for all Ta positions and all Ta atoms  have the same $\nu=\nu_{\rm Q}$. In such a case, the spectra do not split nor shift. Meanwhile, $^{125}$Te with $I=1/2$  is not affected by EFG. Peaks $1$ and $2$ split into two in the CDW state, which can be understood as the result of different Te sites having different Knight shifts in the ordered state.

\subsection{\label{sec:Superconducting}Superconducting state}

\begin{figure}
\includegraphics[width=7cm,clip]{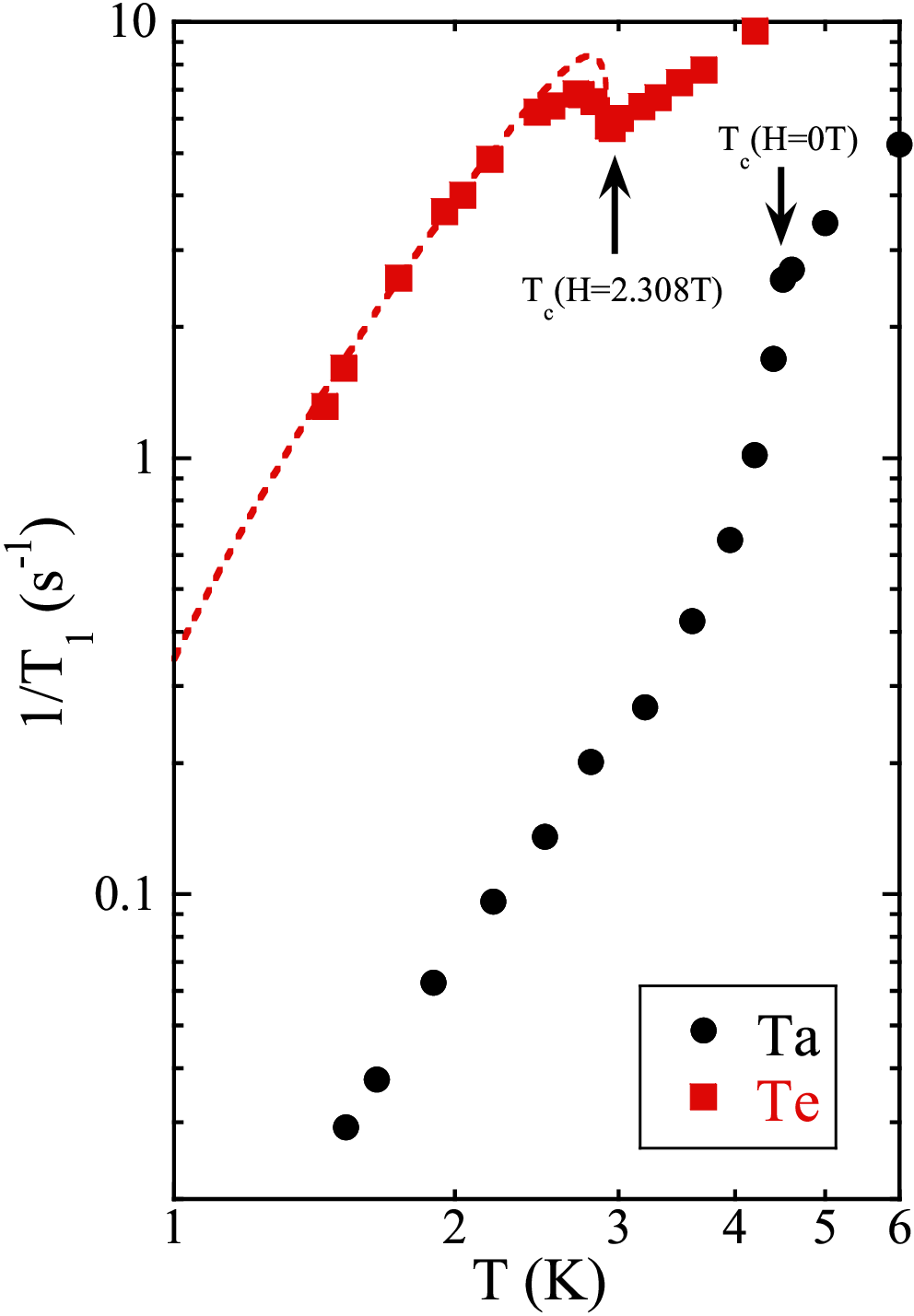}
\caption{\label{fig:Fig_T1} (Color online) The temperature dependence of the spin-lattice relaxation rate of $^{181}$Ta and $^{125}$Te. The curve below $T_{\rm c}$ is a fit to a BCS gap with $\Delta(0)= 1.76k_{\rm B}T_{\rm c}$.}
\end{figure}

The difference in the spin-lattice relaxation rate of $^{125}$Te and $^{181}$Ta is also seen below $T_c$.
Figure \ref{fig:Fig_T1} shows the temperature dependence of $1/T_{1}$ of both $^{125}$Te and $^{181}$Ta.
A coherence peak (Hebel-Slichter peak) is clearly seen for $^{125}$Te, but is absent for $^{181}$Ta.
In most cases, NQR and NMR relaxations is due to the same perturbation, leading to the same spin-lattice relaxation rate. For example, $1/T_{1}T$ from NMR and from NQR show the same behavior in LiFeAs \cite{ZLi2010}.
In the present case, the striking difference  in the $1/T_{1}$ of $^{125}$Te and $^{181}$Ta arises from the fact that $^{125}$Te can only relax via  magnetic interactions while quadrupolar relaxation dominates the $1/T_{1}$ of  $^{181}$Ta at low temperatures.

$1/T_{1}$ in the superconducting state is expressed as
\begin{eqnarray}\label{Eq:T1}
\frac{T_{1{\rm N}} }{T_{1{\rm S}}}=
\frac{2}{k_{\rm B}T N_{0}^{2}}\int \left( 1\mp\frac{\Delta^{2}}{EE'} \right)
N_{\rm S}(E)N_{\rm S}(E')\nonumber\\
f(E)\left[1-f(E')\right]\delta (E-E'){\rm d}E{\rm d}E',
\end{eqnarray}
where $T_{1{\rm N}}$ and $T_{1{\rm S}}$ are the relaxation rates in the normal state and the superconducting state, respectively. $f(E)$ is the Fermi distribution function, and $N_{\rm S}(E)$ is the DOS in the superconducting state. %
$N_{\rm S}(E)=\frac{N_{0}E}{\sqrt{E^{2}-\Delta^{2}}}$, and $C=\left( 1\mp\frac{\Delta^{2}}{EE'} \right)$ is the coherence factor. The sign in $C$ depends on the nature of the interaction that causes the relaxation\cite{MacLaughlinDE}. The magnetic relaxation is due to magnetic interactions between the nuclear spins and electron spins/orbitals, which is not time reversal invariant. In this case, $C=\left( 1\ + \frac{\Delta^{2}}{EE'} \right)$ applies. The relaxation rate has the form
\begin{eqnarray}\label{Eq:T1mag}
\frac{T_{1{\rm N}} }{T_{1{\rm S}}}=
\frac{2}{k_{\rm B}T}\int \left( \frac{E^{2}+\Delta^{2}}{E^{2}-\Delta^{2}} \right)
f(E)\left[1-f(E)\right]{\rm d}E.
\end{eqnarray}
For an $s$-wave gap, the divergence of the DOS at $E=\Delta$ will lead to a Hebel-Slichter peak just below $T_{\rm c}$.
By contrast, the quadrupolar relaxation is due to the electrical interaction between the nuclear quadrupole moment and the EFG, which is time-reversal invariant. In this case, $C=\left( 1-\frac{\Delta^{2}}{EE'} \right)$ applies\cite{HammondRH1960}. The relaxation rate has a simple form
\begin{eqnarray}\label{Eq:T1Q}
\frac{T_{1{\rm N}} }{T_{1{\rm S}}}=
\frac{2}{k_{\rm B}T}\int
f(E)\left[1-f(E)\right]{\rm d}E,
\end{eqnarray}
so $1/T_{1}$ will not show a Hebel-Slichter peak. This explains the difference in the $1/T_1$ results of $^{125}$Te and $^{181}$Ta.
A similar situation was encountered a long time ago in Ta$_{3}$Sn where a coherence peak was seen  for $^{119}$Sn  but quadrupolar relaxation eliminated the coherence peak in the $1/T_1$
 of $^{181}$Ta\cite{Wada1973}.

The  clear Hebel-Slichter coherence peak for $^{125}$Te as seen in Fig. \ref{fig:Fig_T1} indicates that the superconducting gap is of $s$-wave symmetry. If the gap function changes sign over the Fermi surface, such as a $d$ or $p$ wave,  the coherence peak will be suppressed.
To fit the data to a BCS gap, 
 we follow Hebel to convolute $N_{S}(E)$ with a broadening function\cite{Hebel1959}, which is approximated with a rectangular function centered at $E$ with a height of $1/2\delta$.   The  curve in Fig. \ref{fig:Fig_T1} is a calculation with $\Delta(0) = 1.76k_{\rm B}T_{\rm c}$ and  $\delta= \Delta (0)/3.5$, where $\Delta(0)$ is the superconducting gap at the zero-temperature limit. On the other hand, the temperature dependence of the $1/T_1$ below $T_c$ for $^{181}$Ta is determined by both the DOS and the reduction of the EFG fluctuations, which is more complex. A fitting of the data to a theoretical function is beyond the scope of this paper.

Although the electronic structure calculation has shown that Ta$_{4}$Pd$_{3}$Te$_{16}$ is a multi-band system, the main contribution to the DOS is from Te $p$ orbitals. Therefore, the full gap derived from the $^{125}$Te-NMR is also consistent with the suggestion from the band calculation that Ta$_{4}$Pd$_{3}$Te$_{16}$ should be an $s$-wave superconductor mediated by phonons associated with the Te-Te bonding if there are no spin fluctuations\cite{Singh2014}.

\section{\label{sec:Summary}Summary}
In conclusion, we have performed NMR and NQR studies on Ta$_{4}$Pd$_{3}$Te$_{16}$. A striking difference between the spin-lattice relaxation rate $1/T_{1}$ of $^{125}$Te and $^{181}$Ta was found, which indicates that EFG fluctuations develop below around $80$ K. The EFG fluctuations, which reflects charge fluctuations due to a CDW instability, are enhanced with decreasing temperature until a CDW transition sets in at $T_{\rm CDW}=20$ K, below which the resonance frequency, FWHM, and $1/T_{1}$ show sudden changes. In the superconducting state, $1/T_{1}$ of $^{125}$Te shows a clear Hebel-Slichter coherence peak just below $T_{\rm c}$, which is the hallmark of the $s$-wave full gap superconductor.

\begin{acknowledgments}
We thank Q. Ma for assistance in some of the measurements. This work  was supported by the Strategic Priority Research Program of the Chinese Academy of Sciences No. XDB07020200, the Ministry of Science and Technology (MOST) of China under Grant No. 2016YFA0300502, and National Basic Research Program of China (973 Program) No. 2015CB921304.
\end{acknowledgments}





\end{document}